\documentclass[aps,prl,twocolumn,floatfix,showpacs]{revtex4}
\usepackage[dvips]{graphicx}
\usepackage{array}
\usepackage{bm,amssymb,amsmath}

\usepackage{epsf}

\begin{document}


\title{Two-site entropy and quantum phase transitions in low-dimensional models}

\author{\"O.~Legeza and J.~S{\'o}lyom}

\affiliation{Research Institute for Solid State Physics and Optics, H-1525
Budapest, P.\ O.\ Box 49, Hungary }

\date{\today}

\begin{abstract}
We propose a new approach to study quantum phase transitions in 
low-dimensional lattice models. It is based on studying the von Neumann 
entropy of two neighboring central sites in a long chain. It is demonstrated 
that the procedure works equally well for fermionic and spin models, and the 
two-site entropy is a better indicator of quantum phase transition 
than calculating gaps, order parameters or the single-site entropy. The method
is especially convenient when the density-matrix renormalization-group 
(DMRG) algorithm is used. 
\end{abstract}


\maketitle

The search for ground state and the study of quantum phase transitions (QPTs) 
is a challenging problem when strongly correlated fermionic or spin systems are 
considered. Since exactly solvable models are rare, in most cases the relevant part
of the excitation spectrum, the order parameters characterizing the various phases, 
or eventually susceptibilities are determined numerically on finite chains and their 
thermodynamic limit is determined using the standard finite-size scaling method. 
Unfortunately in several cases no definite conclusions can be drawn even if the 
calculations are performed on rather long chains. 

In this letter, we propose a new approach to detect QPTs and to locate the 
quantum critical point in low-dimensional spin or fermionic models. It is based 
on studying the behavior of the von Neumann entropy of two neighboring 
sites in the middle of a long chain, which can be defined both for fermionic and 
spin models, and can be especially easily implemented when the 
density-matrix renormalization-group (DMRG) algorithm \cite{white} is used.  

The method is closely related to concepts in quantum information theory,
which recently have attracted great attention in relation to QPTs. 
Wu \emph{et al.}~\cite{wu} have shown that quite generally QPTs are 
signalled by a discontinuity in some measure of entanglement in the quantum 
system. One such measure is the concurrence~\cite{wootters} which has been 
used by a number of authors~\cite{osborne,osterloh,sylju,gu,vidal,roscilde,yang} 
in their study of spin models. Since the concurrence is defined for spin-1/2 systems 
only, for higher spins or fermionic models another measure 
of entanglement is needed.

\begin{figure}[htb]
\includegraphics[scale=0.6]{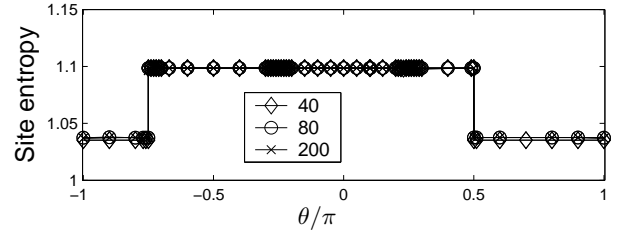}
\vspace{-1mm}\hspace{5cm} $\theta/\pi$ \hfill\\
 \vspace{-3mm}
\caption{$\theta$ dependence of the single-site entropy of the central site of the 
isotropic spin-1 chain, determined for different chain lengths.}
\label{fig:biqu-one-site}
\end{figure}

The local measure of entanglement, the one-site entropy, which is obtained
from the reduced density matrix $\rho_i$ at site $i$, has been proposed by
Zanardi \cite{zanardi} and Gu \emph{et al.} \cite{gu2} to identify QPTs. 
Contrary to their expectation, in many cases, this quantity turns out to be 
insensitive to QPT. As an example let us consider the most general isotropic 
spin-1 chain model described by the Hamiltonian 
\begin{equation}  
        {\cal H} =  \sum_i \big[ \cos\theta ( \bm{S}_i \cdot \bm{S}_{i+1}) +
        \sin\theta ( \bm{S}_i \cdot \bm{S}_{i+1})^2 \big] \,.
\label{eq:ham_biqu}
\end{equation}
In 1D, the model can be solved exactly at $\theta= \pm \pi/4$ and $\theta = \pm 3\pi/4$, 
and is known to have at least four different phases \cite{biqu}. The ground
state is ferromagnetic for $\theta < - 3\pi/4$ and $\theta > \pi/2$, while in between
the integrable points separate the Haldane phase, which exists in the range
$-\pi/4 < \theta < \pi/4$, from the dimerized and the quantum spin nematic 
phases, respectively. The existence of another phase, the quantum quadrupolar
phase \cite{chubu} near $\theta = -3\pi/4$ is not settled yet \cite{buchta}. These 
phases and the corresponding QPTs do not show up at all in the local entropy. As 
can be seen in Fig.~\ref{fig:biqu-one-site}, the site entropy has the value $s_i = \ln 3$ 
in the whole range $-3\pi/4 < \theta < \pi/2$. This is due to the unbroken 
SU(2) symmetry in the nonpolarized states. In the totally aligned ferromagnetic 
state, finite value is obtained for the single-site entropy, because the 
$S^z_{\rm tot} = 0$ component of this state is considered.

The aim of this letter is to point out that a better indicator is obtained if 
instead of the entropy of a single site, the entropy of two neighboring 
central sites, 
\begin{equation}
     s_{i,i+1} = -{\rm Tr} \rho_{i,i+1} \ln \rho_{i,i+1}
\end{equation}
is studied, where $\rho_{i,i+1}$ is the reduced density matrix of the two sites. 
A related quantity has been used in \cite{rissler} to characterize the interaction 
between states in quantum chemistry. We will demonstrate that if---in order to 
avoid boundary effects---sites in the middle of long chains are considered, i.e., 
for $i= N/2$ and $i=N/2+1$, where $N$ is the number of sites in the chain, 
anomalies are developing in this quantity, and their analysis can be used to detect 
quantum phase transitions in low-dimensional lattice models. This procedure is 
especially convenient when a dimerization transition is studied. In this case, namely, 
the difference of two-site entropies on neighboring sites in the center of the chain, 
\begin{equation}
     D_s(i) = s_{i+1,i+2}-s_{i,i+1}\,,  \qquad i=N/2
\end{equation} 
is finite, but the method works for other types of transitions, as well.

As usual in the DMRG approach, we considered open chains. The numerical 
calculations were performed on finite chains up to 400 lattice sites using the 
dynamic block-state selection (DBSS) approach \cite{legeza02}. The threshold 
value of the quantum information loss $\chi$ has been set to $10^{-8}$. 
Similar anomalies appear in the two-site entropy when periodic boundary
condition is used, except that $D_s$ is meaningless in this case.

\begin{figure}[htb]
\includegraphics[scale=0.6]{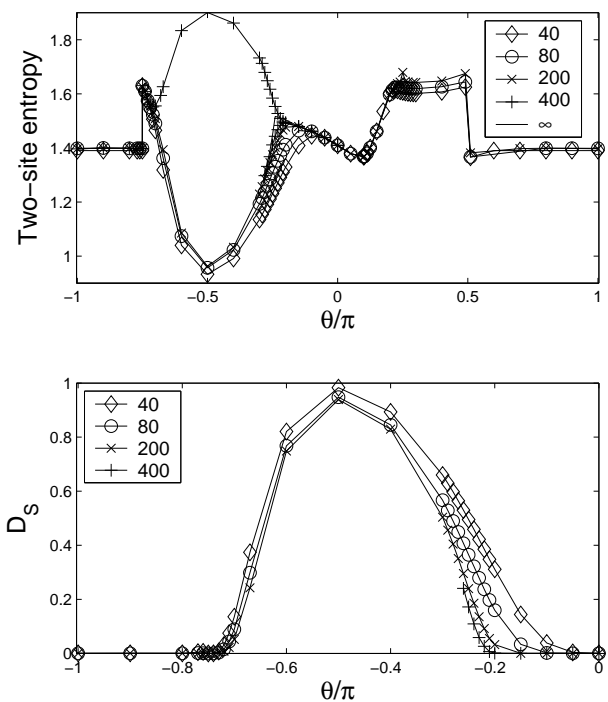}
\vspace{-3mm}
\caption{($a$) $\theta$ dependence of the two-site entropy $s_{i,i+1}$ for $i=N/2$
and $i=N/2+1$ of the isotropic spin-1 chain for different chain lengths. The results for
$i=N/2$ are displayed for $N=400$ only. ($b$) 
Dimerization of the two-site entropy.}
\label{fig:biqu-two-site}
\end{figure}

As a first example the results obtained for the biliniear-biquadratic model are 
shown in Fig. \ref{fig:biqu-two-site}. The upper panel shows $s_{i,i+1}$ for 
$i=N/2$ and $i=N/2+1$, while $D_s(N/2)$ is displayed in the lower panel. 

In the range $-3\pi/4 < \theta < \pi/2$, one can clearly distinguish three 
regimes. The most pronounced feature is that the two-site entropy is 
strongly dimerized for $-3\pi/4 < \theta < - \pi/4$, which reflects the 
dimerization of the valence-bond structure in this regime. The quantity 
$s_{i,i+1}$ for $i=N/2$ and $i=N/2+1$ measures the entropy of a pair of 
sites coupled by strong or weak bonds, respectively. 
The lower panel shows that as longer and longer chains are considered, the 
boundaries of the dimerized region scale to the integrable points, $\theta = -\pi/4$
and $-3\pi/4$. 

In the Haldane phase, where all neighbors are coupled predominantly by one 
valence bond, as in the AKLT state \cite{AKLT}, the two-site entropy is 
not alternating, its minimum is at the AKLT point. Right at $\theta = \pi/4$,
where the model has SU(3) symmetry, the two-site entropy is discontinuous,
moreover in the thermodynamic limit, a kink appears between the left and right 
sides of this point, which is a clear indication of the transition from the Haldane 
phase into the quantum spin-nematic phase. 

Having demonstrated the usefulness of the study of the two-site entropy on
a model where the quantum critical points were known, we now use this 
procedure to study the phase diagram of two 1D fermionic models 
proposed to explain the neutral-ionic (N-I) transition first observed in 
organic mixed-stack charge-transfer salts \cite{torr_01}. 

The ionic Hubbard model \cite{hubb-torr} is defined by the Hamiltonian
\begin{equation}   \begin{split}
      {\mathcal H} &= t \sum_{i\sigma} \left ( c^\dagger_{i\sigma} 
    c^{\phantom\dagger}_{i+1\sigma} + c^\dagger_{i+1\sigma} 
     c^{\phantom\dagger}_{i\sigma} \right )   \\ 
        &  \phantom{=,} + U \sum_{i} n_{i\uparrow} n_{i\downarrow} 
       + \frac{\Delta}{2}\sum_{i\sigma}(-1)^i n_{i\sigma}\,.
\label{eq:ham_ihm}
\end{split}   \end{equation}

When the number of electrons is exactly equal to the number of sites, the 
competition between the on-site energy difference $\Delta$ and the Coulomb 
energy $U$ will determine whether the system is a band insulator $(U < \Delta)$ 
where both the charge and spin gaps are finite or a correlated Mott insulator 
$(U > \Delta)$ where only the charge gap is finite. When hopping is negligibly 
weak, the transition takes place at $U = \Delta$. As was first pointed out 
by Fabrizio \emph{et al}.\ \cite{fabri}, for finite $t$ values the transition 
between these two states occurs in two steps. The charge gap closes and 
reopens at a critical value $U_{\rm c}$, while the spin gap vanishes at a 
different value $U_{\rm s}$. The transition at $U=U_{\rm c}$ is Ising-like 
while the one at $U_{\rm s}$ is a Kosterlitz-Thouless transition. A dimerized 
bond-order phase exists between the two critical values 
$U_{\rm c} < U < U_{\rm s}$. 

\begin{figure}[htb]
\includegraphics[scale=0.6]{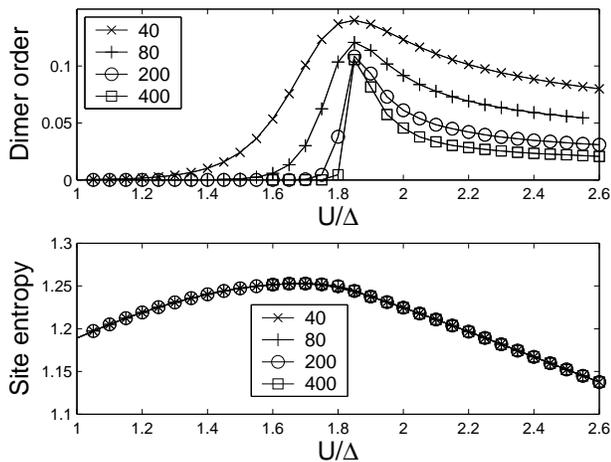}
\vspace{-3mm}
\caption{$(a)$ Dimer order parameter of the ionic Hubbard model for $t/\Delta=1$ 
for different chain lengths. $(b)$ Entropy of the central site.}
\label{fig:dimer-IH}
\end{figure}

Since then the model has been studied in detail by several groups 
\cite{kampf,manma,soos,aligia,leo,otsuka} using both analytic and numerical 
methods. Despite the many efforts the controversy concerning the number of
transitions was not resolved until recently \cite{manma,leo,otsuka}. The 
difficulty can be seen from Fig.\ \ref{fig:dimer-IH}, which shows the dimer order 
parameter (the energy difference of neighboring bonds) for finite chains with 
open boundary condition. Using the standard finite-size scaling procedure the 
vanishing of the dimer order cannot be established with certainty from these data.

\begin{figure}[htb]
\includegraphics[scale=0.6]{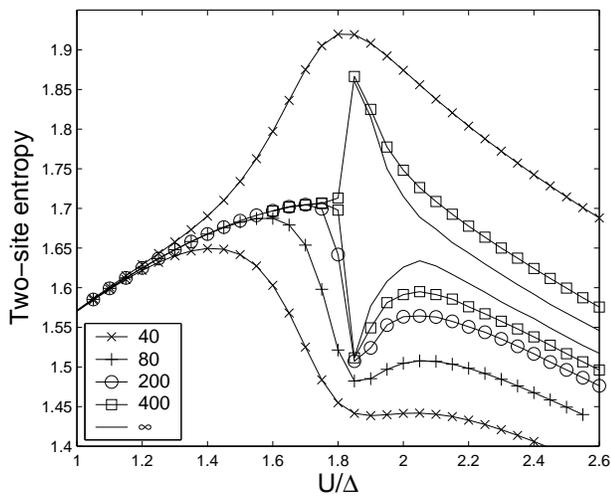}
\vspace{-3mm}
\caption{Two-site entropy $s_{i,i+1}$  for $i=N/2$ and $i=N/2+1$ of the ionic 
Hubbard model for $t/\Delta=1$ for different chain lengths.}
\label{fig:two-IH}
\end{figure}

Moreover, in contrast to what has been found in the extended Hubbard 
model \cite{gu}, the single-site entropy is a rather smooth curve without sharp 
extremum or discontinuity in the derivative. If, however, the two-site entropy 
is analysed, two transitions can be identified. The results for $t/\Delta = 1$ are
shown in Fig. \ref{fig:two-IH}. The two curves for fixed $N$ correspond to
$s_{i,i+1}$ with $i=N/2$ and $i=N/2+1$. Similar behavior has been found for other 
values of $t/\Delta$, too.   

The Ising-like transition, the point where the two-site entropy becomes dimerized,
can be rather well located. On the other hand the Kosterlitz-Thouless transition at 
larger values of the Coulomb coupling, where the dimerization disappears again, 
cannot be established with certainty by looking at $D_s$ alone. Standard finite-size 
scaling does not give a vanishing $D_s$. However, the maxima in $s_{i,i+1}$ for 
$i=N/2 +1$ develop into a cusp in the thermodynamic limit, and this should be 
attributed to the second transition. 

As a next example we have considered another model proposed to
describe the neutral-ionic transition \cite{horovitz}. In this model
donor molecules with ionization energy $I$ and acceptor molecules with 
electron affinity $A$, respectively alternate. If the energies are measured with
respect to the neutral state, in which the donors are doubly occupied and the
acceptors are empty, the molecular levels are described by the
Hamiltonian
\begin{eqnarray} 
\label{eq:hd}
{\mathcal H}_0 &=& I \sum_{i} (2-n_{2i}) +
      U_{\rm D}\sum_i (1-n_{2i,\uparrow})(1-n_{2i,\downarrow})  
     \nonumber   \\
         &  & - A \sum_{i} n_{2i-1} +
          U_A \sum_i n_{2i-1,\uparrow} n_{2i-1,\downarrow}\,,  
\end{eqnarray}
and the hopping, the charge transfer between neighboring donors 
and acceptors is given by
\begin{equation}  \begin{split}
       {\mathcal H}_{\rm CT}&  =  t \sum_{i  \sigma} \big[ c^\dagger_{2i\sigma}
    (c^{\phantom\dagger}_{2i-1,\sigma} + c^{\phantom \dagger}_{2i+1, \sigma}) \\
      & \phantom{=,} +   (c^{\dagger}_{2i-1,\sigma} + c^{\dagger}_{2i+1, \sigma})
         c^{\phantom\dagger}_{2i,\sigma} \big] \,.
\end{split}     \end{equation}

In realistic charge-transfer salts the on-site Coulomb energy is presumably 
the largest energy, so it is reasonable to assume that its unique role is to forbid 
doubly ionized (empty) donors and doubly ionized (doubly occupied) acceptors.
The transition between the neutral and ionic phases is driven by the
intersite Coulomb coupling,
\begin{equation}
      {\mathcal H}_{\rm C} = - V \sum_{i} (2-n_{2i}) (n_{2i-1} +
           n_{2i+1} ) \,.
\end{equation}
The relevant parameter is $V/\Delta$, where $\Delta = \frac{1}{2}(I-A)$.

\begin{figure}[htb]
\includegraphics[scale=0.6]{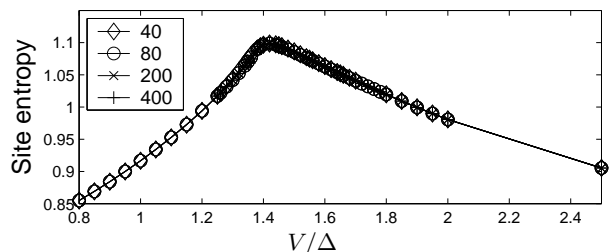}
\vspace{-1mm}\hspace{5cm} $V/\Delta$ \hfill \\
\vspace{-3mm}
\caption{Single-site entropy of the donor-acceptor model for $t/\Delta=1$ for 
different chain lengths.}
\label{fig:t1_1site}
\end{figure}

This problem has been studied by exact diagonalization on relatively 
short chains \cite{horovitz} as well as by valence-bond techniques \cite{paine}. 
For small transfer integral the model was found  to exhibit a first-order 
phase transition form a neutral to an ionic phase as a function of $V/\Delta$. 
For $t/\Delta$ larger than $0.3$ the transition changes character 
and a second-order phase transition has been reported. Our numerical 
result for the site entropy for $t/\Delta =1$ is shown in Fig.~\ref{fig:t1_1site}. 
A cusp is clearly seen in the one-site entropy at $V/\Delta \simeq 1.42$, indicating 
a single second-order transition, in agreement with earlier expectations. 

\begin{figure}[htb]
\includegraphics[scale=0.6]{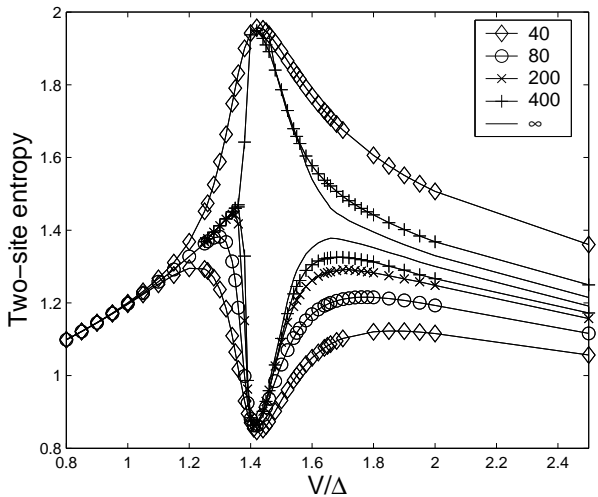}
\vspace{-3mm}
\caption{Two-site entropy of the donor-acceptor model for $t/\Delta=1$ for 
different chain lengths.}
\label{fig:t1_2site}
\end{figure}

Contrary to this the two-site entropy shown in Fig.~\ref{fig:t1_2site} exhibits 
the same behavior as the ionic Hubbard model. In fact, as will be shown 
in a separate paper, the two models are limiting cases of a unified model of 
neutral-ionic transition. Although, as for the ionic Hubbard model, it is not easy to 
locate the vanishing of dimerization when $D_s$ alone is considered, the peak 
of $s_{i,i+1}$ for $i=N/2+1$ produces a cusp in the thermodynamic limit, indicating 
a second transition. For small transfer integral, $t/\Delta<0.3$, the two transition 
points coalesce, both the one and two-site entropies exhibit a sharp jump, the 
transition is of first order. 

In conclusion, we have shown that the entropy of two central sites in a long open 
spin or fermionic chain, which can be easily calculated in the density-matrix 
renormalization-group procedure provides us with extra information not contained 
in the single-site entropy. The two-site entropy displays maxima or break points 
at QPT even in such cases when it is difficult to establish the opening or closing 
of a gap, or the vanishing of an order parameter, or when the single-site entropy is 
featureless since it is insensitive to the breaking of symmetry that distinguishes the 
two phases. Using this procedure we have demonstrated that the ionic Hubbard 
model and the donor-acceptor model do in fact have two phase transitions. 

\begin{figure}[htb]
\includegraphics[scale=0.6]{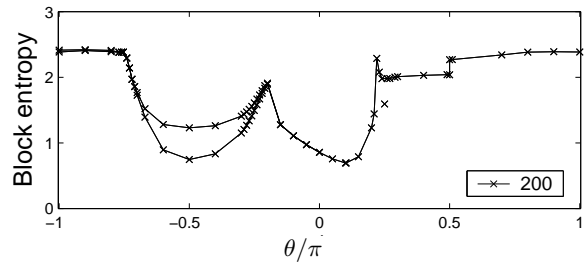}
\vspace{-1mm}\hspace{5cm} $\theta/\pi$ \hfill \\
\vspace{-3mm}
\caption{Entropy of blocks of $N/2$ and $N/2+1$ sites of the bilinear-biquadratic 
$S=1$ model for a chain with $N=200$ sites.}
\label{fig:block}
\end{figure}

The procedure can be easily extended to considering the entropy of not just
two-sites, but that of a longer segment, whose length dependence is different
whether the model is critical or not \cite{vid}. Fig.~\ref{fig:block} shows
the entropy of the block of $N/2$ and $N/2 + 1$ sites of an $N=200$ chain 
for the bilinear-biquadratic $S=1$ model. The dimerization appears in
the block entropy, as well, and the anomalies whose location scales to the 
transition points $ \theta = \pm \pi/4$ are more pronounced than in the two-site 
entropy. We have learned, after the submission of our paper, that the study of 
the block-block entanglement in connection with quantum phase transition has 
been proposed by Deng \emph{et al.}\ \cite{deng}.  Our experience shows, 
however, that calculating the block entropy with the same accuracy as that of 
two sites is a numerically much more demanding task.

This research was supported in part by the Hungarian Research Fund (OTKA)
Grants No.\ T043330 and F046356. The authors acknowledge
computational support from Dynaflex Ltd under Grant No. IgB-32.
\"O. L. was also supported by the J\'anos Bolyai scholarship.


\begin{thebibliography}{99}

\bibitem{white}
S.R. White, Phys. Rev. Lett. {\bf 69},  2863  (1992); Phys. Rev. B
         {\bf 48},  10345  (1993).

\bibitem{wu}
L.-A. Wu, M.S. Sarandy, and D.A. Lidar, Phys. Rev. Lett. {\bf 93}, 
250404 (2004). 

\bibitem{wootters}
W.K. Wootters, Phys. Rev. Lett. {\bf 80}, 2245 (1998). 

\bibitem{osborne}
T.J. Osborne and M.A. Nielsen, Phys. Rev. A {\bf 66}, 32110 (2002).

\bibitem{osterloh}
A. Osterloh, L. Amico, G. Falci, and R. Fazio, Nature {\bf 416}, 608 (2002).  

\bibitem{sylju}
O.F. Sylju{\aa}sen, Phys. Rev. A {\bf 68}, 60301(R) (2003). 

\bibitem{gu}
S.-J. Gu, H.-Q. Lin, and Y.-Q. Li, Phys. Rev. A {\bf 68}, 42330 (2003). 

\bibitem{vidal}
J. Vidal, G. Palacios, and R. Mosseri, Phys. Rev. A {\bf 69}, 022107 (2004); 
J. Vidal, R. Mosseri, and J. Dukelsky, {\sl ibid.} {\bf 69}, 054101 (2004). 

\bibitem{roscilde}
T. Roscilde, P. Verrucchi, A. Fubini, S. Haas, and V. Tognetti, Phys. Rev. 
     Lett. {\bf 93}, 167203 (2004); {\bf 94}, 147208 (2005). 

\bibitem{yang}
M.-F. Yang, Phys. Rev. A. {\bf 71}, 30302(R) (2005). 

\bibitem{zanardi}
P. Zanardi, Phys. Rev. A {\bf 65}, 42101 (2002).

\bibitem{gu2}
S.-J. Gu, S.-S. Deng, Y.-Q. Li, and H.-Q. Lin, Phys. Rev. Lett. {\bf 93}, 
86402 (2004). 

\bibitem{biqu}
For its phase diagram see G. F\'ath and J. S\'olyom, Phys. Rev. B {\bf 44},  
11836  (1991);  {\bf 47},  872  (1993); {\bf 51},  3620  (1995).

\bibitem{chubu}
A.V. Chubukov, J. Phys. Condens. Matter {\bf 2},  1593  (1990);
     Phys. Rev. B {\bf 43},  3337  (1991); A. L{\"a}uchli, G. Schmid, 
    and T. Trebst,  cond-mat/0311082 (2003).

\bibitem{buchta}
K. Buchta, G. F\'ath, {\"O}. Legeza and J. S\'olyom, Phys. Rev. B {\bf 72},  
     054433 (2005).

\bibitem{rissler}
J. Rissler, R.M. Noack, and S.R. White, arXiv:cond-mat/0508524 (2005).

\bibitem{legeza02}
{\"O}. Legeza, J. R{\"o}der, and B.A. Hess, Phys. Rev. B {\bf 67},  125114
  (2003); {\"O}. Legeza and J. S\'olyom, Phys. Rev. B {\bf 70},  205118  (2004).

\bibitem{AKLT}
I. Affleck, T. Kennedy, E. Lieb, and H. Tasaki, Phys. Rev. Lett. {\bf 59}, 
799 (1987). 

\bibitem{torr_01}
J.B. Torrance, J.E. Vazquez, J.J. Mayerle, and V.Y. Lee, Phys. Rev. Lett. 
{\bf 46}, 253 (1981).

\bibitem{hubb-torr}
J. Hubbard and J.B. Torrance, Phys. Rev. Lett {\bf 47}, 1750 (1981).

\bibitem{fabri}
M. Fabrizio, A.O. Gogolin, and A.A. Nersesyan, Phys. Rev. Lett. {\bf 83}, 
2014 (1999); Nucl. Phys. B {\bf 580}, 647 (2000).

\bibitem{kampf}
A.P. Kampf, M. Sekania, G.I. Japaridze, and P. Brune, J. Phys.: Condens. 
Matter {\bf 15}, 5895 (2003).

\bibitem{manma}
S.R. Manmana, V. Meden, R.M. Noack, and K. Sch\"onhammer, Phys. Rev. 
    B {\bf 70}, 155115 (2004).

\bibitem{soos}
Z.G. Soos, S.A. Bewick, A. Peri, and A. Painelli, J. Chem. Phys. {\bf 120}, 
    6712 (2004).

\bibitem{aligia}
A.A. Aligia and C.D. Batista, cond-mat/0412026 (2004).

\bibitem{leo}
L. Tincani, R. Noack, and D. Baeriswyl (unpublished). 

\bibitem{otsuka}
H. Otsuka and M. Nakamura, Phys. Rev. B {\bf 71}, 155105 (2005).

\bibitem{horovitz}
B. Horovitz and J. S\'olyom, Phys. Rev. B {\bf 35}, 7081 (1987).

\bibitem{paine}
A. Painelli and  A. Girlando, Phys. Rev. B {\bf 37}, 5748 (1988).

\bibitem{vid}
G. Vidal, J.I. Latorre, E. Rico, and A. Kitaev, Phys. Rev. Lett. {\bf 90}, 227902 (2003).

\bibitem{deng}
S.-S. Deng, S.-J. Gu, and H.-Q. Lin, arXiv:quant-ph/0511103 (2005). 

\end{thebibliography}
\end{document}